\documentstyle[prl,twocolumn,epsfig,aps]{revtex}
\tighten
\begin{document}
\draft

\twocolumn[\hsize\textwidth\columnwidth\hsize\csname @twocolumnfalse\endcsname
\title{Soliton back-action evading measurement using spectral filtering}
\author{F. K\"onig, B. Buchler$^*$, T. Rechtenwald, G.~Leuchs and A. Sizmann}
\address{Lehrstuhl f\"ur Optik, Physikalisches Institut der Universit\"at Erlangen-N\"urnberg, \\ Staudtstr. 7/B2, 91058 Erlangen, Germany}
\address{$^*$permanent address: Department of Physics, Faculty of Science,\\
The Australian National University, A.C.T.0200, Australia}
\date{\today}
\maketitle
\begin{abstract}
We report on a back-action evading (BAE) measurement of the photon
number of fiber optical solitons operating in the quantum regime.
We employ a novel detection scheme based on spectral filtering of
colliding optical solitons. The measurements of the BAE criteria
demonstrate significant quantum state preparation and transfer of
the input signal
 to the signal and probe outputs exiting the apparatus,
displaying the quantum-nondemolition (QND) behavior of the
experiment.

\end{abstract}
 ]
\section{Introduction}
Quantum theory allows for the measurement of an observable with
arbitrarily high precision. In a back-action evading (BAE)
measurement the back action inherent in the process of a quantum
measurement is confined to the observable which is conjugate to
the measured one \cite{BraVor80}. If the initial state of a system
is an eigenstate of the measurement operator, it is conserved and
the BAE interaction is called quantum-nondemolition (QND).
Initially designed for the detection of gravitational waves, the
concept of QND measurements now is finding applications in the
field of quantum information processing and communication
\cite{LeuSil01}.

Most BAE measurements have been carried out in quantum optics,
where the quantum fluctuations are readily measurable.
Quantitative criteria for QND and BAE experiments have been
developed \cite{ImoSai89,HolCol90,PoiRoc94}.  There are three
physical systems in which BAE detection has been pursued
experimentally: The $\chi^{(3)}$ interactions in optical fibers
\cite{LevShe86,FriMac92} and in atoms \cite{GraRoc91}, and the
$\chi^{(2)}$ interaction in crystals \cite{LaPSlu89}. The QND
criteria were fulfilled in BAE detection with atomic systems
($\chi^{(3)}$) \cite{PoiGra93,RocVig97} and $\chi^{(2)}$
interactions in crystals
\cite{LevAbr93,PerOu94,SchBru96,BruSch97a,BucAnd01}. Repeated BAE
measurements were reported using OPAs \cite{BenLev95,BruHan97} and
a demonstration of principle using fiber optical
pulses\cite{FriMuk00}. Despite their large potential for QND
measurements \cite{HauWat89,DruBre94,CouSpa98,KozIva02} fiber
optical experiments have not previously been performed in the BAE
regime \cite{GraLev98,FriMac92}.

In this Paper we report experimental BAE results following a
recently proposed fiber-optical back-action evading measurement
scheme of the soliton photon number using spectral filtering of
solitons \cite{KonZie02}.

Optical solitons are a natural physical system for implementing
quantum communication concepts in fibers. Recently, they were used
for EPR-pair beam generation \cite{SilLam01}. Solitons are fiber
optical pulses maintaining a stable shape due to a balance of
group-velocity dispersion and nonlinearity. Moreover, after the
collision of two solitons, the pulses recover energy, velocity and
shape, and thus behave like particles. The nonstationary evolution
of the soliton quantum fluctuations leads to remarkable effects.
Quadrature squeezing was the first effect in a series of quantum
optics experiments with solitons in fibers
\cite{RosShe91,SizLeu99}. In recent years amplitude squeezing of
solitons using spectral filtering was discovered and explained as
a multi mode quantum effect \cite{Wer96,FriMac96,SpaKor98}.

The technique of spectral filtering has since been generalized to
a system of two colliding solitons \cite{KonZie02}. A new
theoretical proposal for BAE measurements was derived, in which a
"signal" soliton collides with a second "probe" soliton. During
the collision, the frequency of the probe becomes coupled to the
photon number of the signal. A measurement of the probe frequency
therefore gives a BAE readout of the signal photon number. Using
this scheme we have obtained experimental BAE operation in the
quantum regime. For the first time in fiber QND detection, the
standard QND criteria are clearly fulfilled.

\section{Principle of the BAE measurement}

\begin{figure}
\begin{center}
\epsfig{file=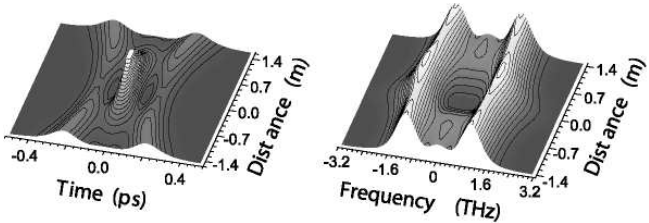,width=8.6cm} \vspace{-0.1cm}
\caption{\label{1)collision} Evolution of the temporal and
spectral power densities of two colliding solitons. Their initial
frequency separation ($=$ relative velocity) in soliton units is
$2\Omega_0 = 2.4 $. The center of collision is at a distance $\xi
= 0$. The soliton period is $1.4$ m and in soliton units is given
by: $\xi_0 = \pi/2$.}
\end{center}
\end{figure}

The classical field evolution in a single-mode polarization
preserving fiber is governed by the nonlinear Schr\"odinger
equation (NLSE):
\begin{equation}
 i\frac{\partial  u}{\partial \xi}+
\frac{1}{2}\frac{\partial^2  u}{\partial \tau^2} +  |u|^2  u= 0 , \label{nlse}
\end{equation}
where $u$ is the slowly varying envelope of the electric field and
$\xi$ and $\tau$ are time and space variables in a reference frame
moving with the optical field \cite{Agr95}. $u$, $\xi$ and $\tau$
are given in soliton units \cite{MolEva91}. The single pulse
soliton solutions, i.e. the fundamental soliton of amplitude $A$
is given by \cite{MolEva91}:
\begin{eqnarray}u(\tau, \xi) = A\,
{\rm sech}(A(\tau+\Omega\xi ))\exp(i (A^2-\Omega^2) \xi/2 - i
\Omega \tau). \nonumber \end{eqnarray}
$\Omega$ is a dimensionless velocity of the soliton in units of
inverse pulse length relative to the moving frame. At the same
time $\Omega$ is a dimensionless center frequency of the soliton,
because of the group-velocity dispersion. The collision of two
fundamental solitons of equal amplitudes $A$, moving before the
collision with $\pm \Omega_0$, is depicted in Figure
\ref{1)collision}.
\vspace{-0.0cm}
\vspace{-0.4cm}
As the solitons begin to overlap the enhanced intensity in the
overlap region causes an attractive force and acceleration between
the pulses. Therefore, the initial spectral separation and
relative velocity of the solitons is transiently enhanced above
the initial value of $2\Omega_0$. The spectral separation of the
pulses, "signal" and "probe", changes to $2\Omega_0 + d\Omega_S +
d\Omega_P$ and depends on the photon number of both solitons.
Therefore, if the signal amplitude $A_S$ fluctuates around $A =
\langle A_S \rangle$, the fluctuations $ \Delta A_S = A_S - A $
induce fluctuations in the frequency $\Delta\Omega_P $ of the
probe soliton at the collision center. If the soliton spectra do
not overlap ($\Omega_0\gg 1$), the probe shift is given by
\cite{Formel}:
\begin{equation}
\langle d\Omega_P\rangle + \Delta\Omega_P = \frac{2 A^2}{3
\Omega_0} (1+\frac{\Delta A_S}{2 A}+{\mathcal O}(\Delta A_S^2))
\label{frequencyshift}
\end{equation}
Towards the end of the collision the solitons recover their initial velocities
(Fig.\ref{1)collision}).

The main idea of the novel BAE scheme is to measure the signal
soliton amplitude fluctuations via the probe soliton frequency.
The coupling is described classically by Eq. \ref{frequencyshift}.
Signal and probe solitons can be spectrally separated at the
collision center (Fig.\ref{1)collision}). Then the frequency
fluctuations $\Delta\Omega_P $ of the probe are read out with a
spectral edge filter. The signal, which is the photon number, is
preserved in the interaction as well as in the free fiber
propagation, because of negligible fiber loss. The back action of
the measurement perturbs the frequency and the phase of the signal
pulse. For a quantum readout, the probe readout has to determine
the signal intensity to better than the signal shot noise
uncertainty. This measurement scheme was assessed by a detailed
theoretical investigation accounting for the quantization of the
field $u$. The result indicates that the signal is measured more
precisely than within the shot noise and the scheme fulfills the
QND criteria \cite{KonZie02}.

\section{Experimental realization}
\begin{figure}
\begin{center}
\epsfig{file=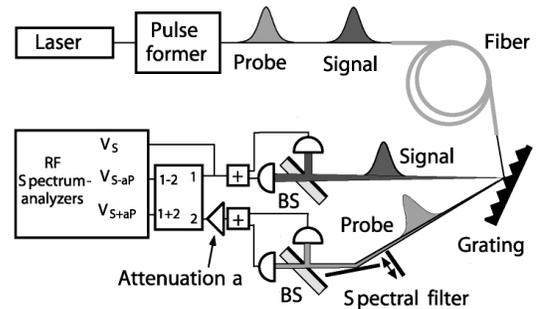,width=7.0cm} \vspace{0.0cm}
\caption{\label{2)setup} BAE apparatus for pulse pair generation,
soliton collision and spectrally filtered probe detection (BS =
beam splitter with $50\%$ reflection/transmission).}
\end{center}
\end{figure}

\begin{figure}
\begin{center}
\epsfig{file=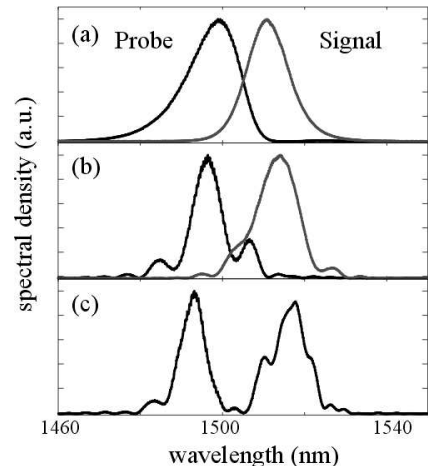,width=5.5cm} \vspace{0.1cm}
\caption{\label{3)spektra} Signal and probe spectra out of the
pulse former (a), after separate fiber propagation (b), and of a
pulse collision (c) (relative phase of $\pi$).}
\end{center}
\end{figure}

The experimental realization of the measurement requires three
main steps: the pulse pair for the soliton collision has to be
prepared and launched into the fiber. Then the detector is set up
for the measurement of the conditional variance. Finally, the
transfer coefficients are measured using an RF modulation on the
signal input soliton. An overview of the experimental apparatus is
presented in Fig. \ref{2)setup}. The laser source is a $\rm
{Cr}^{4+}$:YAG laser delivering 150-fs ${\rm sech}$-pulses at
$\lambda = 1.5 \mu$m \cite{SpaBoh97}. In the pulse former every
pulse is spectrally split using a wavelength dependent beam
splitter and retroreflected, double passing the beam splitter such
that two pulses of different center wavelengths and with an
adjustable temporal separation are obtained. They exhibit no
relative phase noise or timing jitter in contrast to pulses from
different laser sources. Figure \ref{3)spektra}(a) displays the
two pulses after the pulse former. They have a spectral width of
12~nm and are separated by 14~nm.
%
\vspace{-0.5cm}
\vspace{-0.4cm}
The temporal width of the autocorrelation traces indicates pulse
lengths of 200~fs. The time-bandwidth product is slightly
increased above the Fourier limit (sech-shape assumed). In
consequence, a weak dispersive wave evolves in the fiber in
addition to the two solitons. The timing of the input pulses was
adjusted to the output wave autocorrelation trace, taken for every
measurement, such that the fiber end is exactly in the middle of
the collision ($\xi = 0$). As the signal to be measured in the BAE
apparatus we consider the amplitude of the signal pulse after it
is coupled into the fiber; therefore the coupling losses do not
degrade the signal transfer through the BAE apparatus. Before the
fiber the input pulse pair was shown to be shot-noise limited
around 20 MHz to within $\pm 0.2$ dB. We used 6.3 m of
polarization preserving single mode fiber (3M, FS-PM 7811)
corresponding to 3.2 and 2.9 soliton periods \cite{Agr95} for the
signal and probe solitons, respectively. The collision length,
i.e. the length of the fiber where the solitons overlap within
their (power) half width, is $L_{\rm coll}= 3.2$ m. The solitons
experience a nonlinear cross-phase shift of $\delta \phi_{NL}
\approx 1.0$ rad. An output spectrum is shown in Fig.
\ref{3)spektra}(c) and can be compared to the spectra of pulses
which have propagated individually without collision (3(b)). The
strong nonlinear coupling is clearly observable. The respective
spectra shift about half a spectral width, $d \lambda \approx 5$
nm. The undulations in the spectra are caused by the weak
dispersive wave. Due to the partial spectral overlap ($\Omega_0 =
1.2$) the soliton collision is phase dependent. This modifies the
undulations, but the spectral shift changes by less than $5\%$.

The output pulses are spectrally dispersed using a grating (600
l/mm). Spectral filtering of the probe pulse is accomplished with
a knife edge filter in the Fourier plane of the grating. Signal
and probe pulses are separately directed onto two balanced
two-port detectors (Epitaxx ETX500 photodiodes). The RF
photocurrents are amplified after suppression of the repetition
rate of the laser (163MHz). The sum current $I_P$ of the probe
detector is attenuated with an attenuation $a$ and added to and
subtracted from the signal sum current $I_S$. The shot-noise
levels were determined by reading the DC photocurrents which
display the average detected powers. They were carefully and
repeatedly calibrated to the shot-noise level using the difference
photocurrents of the balanced detectors which carry fluctuations
of the signal and probe shot-noise levels.

The overall quantum efficiency for signal and probe detectors,
including photodiode efficiencies, was measured to be 74.5\% and
69.8\%, respectively. The fiber loss ($\alpha < 0.002{\rm dB/m}$)
was neglected. Coupling losses from the fiber were minimized by
means of an antireflection coated gradient-index lens.

The conditional variance and the transfer coefficients quantify
the performance of a BAE experiment
\cite{ImoSai89,HolCol90,PoiRoc94}. The conditional variance is a
measure of the quantum state preparation ability of the apparatus.
It can be written in terms of the photocurrents $I$ as:
\begin{equation}
   V_{S/P} = \min_a\left(\frac{\langle(I_S+a I_P)^2\rangle}
   {\langle I_{S,SN}^2\rangle}\right)
   \label{eqvcondexp}
\end{equation}
assuming $\langle I \rangle = 0$. $I_S$ ($I_P$) measures the
signal (probe) amplitude fluctuations, since the signal (probe) is
directly detected. $I_{S,SN}$ denotes the corresponding shot
noise.

\section{Results}
\begin{figure}
\begin{center}
\epsfig{file=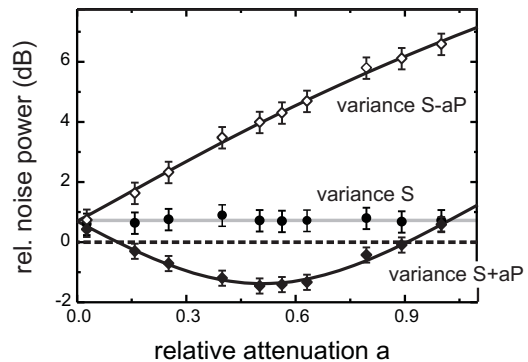,width=6.8cm}
\vspace{0.2cm}\caption{\label{figcond} Photocurrent noise levels
of combined photo currents, normalized to the signal output shot
noise. Note that the combined noise decreases below signal output
shot noise.}
\end{center}
\end{figure}

A measurement of the conditional variance is shown in Fig.
\ref{figcond}. The three noise power levels of signal, sum and
difference currents are recorded simultaneously on three RF
spectrum analyzers. The photocurrent fluctuations are measured at
$20$ MHz with a $300$ kHz resolution bandwidth. Thermal noise in
the detector electronics is $6$ dB below the lowest detected noise
level and is subtracted off in Fig. \ref{figcond}. The noise
powers are quadratic functions of the attenuation $a$ as indicated
by the parabolic fitting curve of Eq. \ref{eqvcondexp}. The
displayed noise levels are normalized to the signal output shot
noise. Signal and probe pulse are separated at $\lambda_{\rm sep}
= 1506$ nm after the fiber; the frequency low pass filter for the
probe was applied at $\lambda_{\rm filt} = 1490$ nm. At this
position roughly $18\%$ of the probe pulse energy is filtered off
and absorbed. The sum noise level in Fig. \ref{figcond} decreases
below the signal noise. This noise suppression corresponds to a
strong negative photon number correlation of $C_{S/P}= -0.62 \pm
0.03$, with the correlation coefficient defined as in
\cite{PoiRoc94,KonZie02}.
\vspace{-0.5cm}
An anti-correlation is anticipated since an increase in the photon
number of the signal causes an {\it enhancement} in the mutual
spectral shift, in turn causing more losses in the probe pulse at
the spectral filter. This correlation is not corrected for the
finite quantum efficiencies of the detection apparatus. The noise
reduction reaches $1.37 \pm 0.3$ dB below the shot noise level of
the signal, corresponding to a conditional variance of
$V_{S/P}=0.73 \pm 0.04$. The necessary condition for a QND
experiment is $V<1$. This is clearly fulfilled.

The second important criterion for a QND measurement is the sum of
the transfer coefficients. They describe the nondemolition
property of the measurement with the transfer of the optical
signal-to-noise ratio from the input to the output:
\begin{eqnarray}
    {\rm T}_S = \frac{{\rm SNR}^{\rm out}_S}{{\rm SNR}^{\rm in}_S} = \frac{\langle(\Delta n_S^{\rm out})^2\rangle\langle n_S^{\rm in}\rangle}{\langle(\Delta
    n_S^{\rm in})^2\rangle\langle n_S^{\rm out}\rangle}\\
    \label{transfer}
    {\rm T}_P = \frac{{\rm SNR}^{\rm out}_P}{{\rm SNR}^{\rm in}_S} = \frac{\langle(\Delta n_P^{\rm out})^2\rangle\langle n_S^{\rm in}\rangle}{\langle(\Delta
n_S^{\rm in})^2\rangle\langle n_P^{\rm out}\rangle},
\end{eqnarray}
where $n$ denotes the photon numbers and $\Delta n$ is the
modulation strength.
Therefore we modulated the intensity of the signal beam with a
$\mbox{LiTaO}_3$-electro-optic modulator at $20$ MHz before
coupling into the fiber. The noise levels on and near to this
frequency ($\sim 50$ kHz off the 20 MHz signal, recorded with a
resolution bandwidth of 10 kHz.) were taken and the
signal-to-noise ratio (${\rm SNR}$) for the signal input, the
signal output and the probe output was determined \cite{PoiGra93}.
If the probe soliton is absent, the unperturbed signal input exits
the fiber (fiber damping and output coupling losses neglected).
This way the signal input ${\rm SNR}$ is measured on the signal
output detector. Therefore, the signal in- and output ${\rm SNR}$s
can be measured with the same two-port detector. The signal
transfer coefficient ${\rm T}_S$ is thus determined independently
of detector efficiencies as the {\it optical} ${\rm SNR}$ transfer
coefficient. The probe ${\rm SNR}$ is measured with the probe
detector. As a consequence, the measured probe ${\rm SNR}$ has to
be corrected for the small difference in signal and probe detector
efficiencies of $6.3\%$. In a test, a weak phase modulation of the
signal showed no ${\rm SNR}$ transfer to the probe. For an
optimized filter, mainly the amplitude fluctuations of the signal
couple to the detected part of the probe, despite the slight phase
sensitivity of the collision. This also shows that the experiment
is unaffected by thermal or quantum phase noise.

The best transfer coefficients measured were ${\rm T}_S + {\rm
T}_P = 0.99 + 0.63 = 1.62 (\pm 0.15)$ at the filter settings of
$\lambda_{\rm sep} = 1504$ nm; $\lambda_{\rm filt} = 1492$ nm.
This was measured with an ${\rm SNR}$ of the incoming signal of
103, almost identical signal-out ${\rm SNR}$ and a probe-output
${\rm SNR}$ of 63. This high value of ${\rm T}_S + {\rm T}_P$
demonstrates the optical tapping property of the device.

\begin{figure}
\begin{center}
\epsfig{file=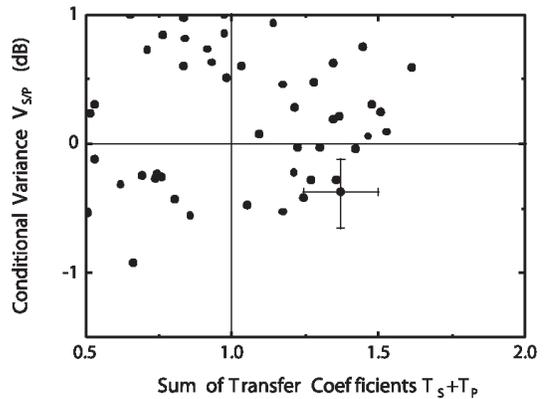,width=7cm}\vspace{0.2cm}
\caption{\label{5)T&V} Data points obtained with the soliton
collision experiment under different parameters, according to
their transfer-coefficients and conditional variance.}
\end{center}
\end{figure}

Simultaneously optimizing the apparatus for both, a low
conditional variance and high transfer coefficients, we observed
$V_{S/P}=0.92 \pm 0.05$ and ${\rm T}_S+{\rm T}_P = 1.37 \pm 0.13$
($\lambda_{\rm sep} = 1503$~nm; $\lambda_{\rm filt} = 1490$~nm).
This result clearly fulfills both the QND criteria. Depending on
the relative phase of signal and probe as well as the filter
positions ($\lambda_{\rm sep}$ and $\lambda_{\rm filt}$) the
achieved transfer coefficients and conditional variances differ.
The data obtained are presented in Fig.\ref{5)T&V}. Stable results
were obtained in all of the four quadrants. To achieve performance
in the QND domain, we separated signal and probe at $1500$~nm$ <
\lambda_{\rm sep} < 1506.5$~nm and low-pass filtered the probe at
$1488$~nm$ < \lambda_{\rm filt} < 1490$~nm. The experimental
results for $V_{S/P}$ and ${\rm T}_S+{\rm T}_P$ are not
correlated. This suggests that additional noise is introduced to
the solitons. Likely candidates are inter- and intra-soliton
stimulated Raman scattering \cite{ChiWen89}. The Raman effect
pumps photons towards the red and depletes the probe. This
assumption is supported by an observed imbalance in the power of
the output pulses of $\Delta P/P \approx \pm 8\%$ and by the
observation of excess noise in the signal output
(Fig.\ref{figcond}). The effect decreases the observed signal
transfer coefficient. A second perturbative effect is the evolving
dispersive wave which modifies the quantum fluctuations
significantly, because spectral correlations not only arise
between the solitons, but also between them and the continuum
\cite{MecKum97,LevVas99,SchKno00,LohGru94}. The use of longer
pulses would reduce the Raman effect; the dispersive wave can be
avoided with improved input pulse shaping.

\vspace{-0.2cm}
\vspace{-0.5cm}
A full QND experiment can be performed with repeated BAE
measurements. When the signal is freely propagated after a first
BAE interaction, the soliton will reshape with only little losses
\cite{KonZie02} and can be used in a second BAE measurement. A way
to nearly eliminate residual radiation loss for cascaded BAE
detection is the insertion of a second probe pulse in the center
of the collision, prepared to complete the collision.

\section{Conclusion}
In conclusion, we demonstrated a back-action evading measurement
of the photon number based on the novel concept of spectral
filtering. This new technique is robust against phase noise
limiting previous experiments \cite{FriMac92}. It requires merely
two pulses since no phase reference pulse is needed and only
direct detection is used for the BAE readout. The experiment
showed large capacity for optical tapping as well as the
capability for quantum state reduction. The idea of coupling to a
completely different degree of freedom in the probe, neither
conjugate nor identical to the signal observable may be utilized
to improve BAE or QND measurements in fibers as well as in other
$\chi^{(3)}$ and $\chi^{(2)}$ systems. Applications can be found,
e. g. in quantum information \cite{BraPat01} or weak absorption
measurements \cite{SouSch97}. Investigating the quantum properties
of soliton collisions explores ultimate bounds in
wavelength-division multiplexed optical transmission systems
\cite{MolEva91,SizKon00}.

\section*{Acknowledgement}
We extend special thanks to Ch. Silberhorn, B. Pfeiffer and M. Meissner for their help in
the data acquisition. We gratefully acknowledge fruitful discussions with P. K. Lam, S.
Sp\"alter and N. Korolkova. This project was supported by the Deutsche
Forschungsgesellschaft.


\begin{thebibliography}{10}

\bibitem{BraVor80}
V.~B. Braginsky, Y.~I. Vorontsov, and K.~S. Thorne, Science {\bf
209},  547
  (1980).

\bibitem{LeuSil01}
G. Leuchs, C. Silberhorn, F. K\"onig, P.~K. Lam, A. Sizmann, and
N. Korolkova,
  in {\em Quantum Information Theory with Continuous Variables}, edited by
  S.~L. Braunstein and A.~K. Pati (Kluwer Academic, Dordrecht, 2002).

\bibitem{ImoSai89}
N. Imoto and S. Saito, Phys. Rev. A {\bf 39},  675  (1989).

\bibitem{HolCol90}
M.~J. Holland, M.~J. Collett, D.~F. Walls, and M.~D. Levenson,
Phys. Rev. A
  {\bf 42},  2995  (1990).

\bibitem{PoiRoc94}
J.~P. Poizat, J.~F. Roch, and P. Grangier, Ann. Phys. Fr. {\bf
19},  2995
  (1994).

\bibitem{LevShe86}
M.~D. Levenson, R.~M. Shelby, M. Reid, and D.~F. Walls, Phys. Rev.
Lett. {\bf
  57},  2473  (1986).

\bibitem{FriMac92}
S.~R. Friberg, S. Machida, and Y. Yamamoto, Phys. Rev. Lett. {\bf
69},  3165
  (1992).

\bibitem{GraRoc91}
P. Grangier, J.~F. Roch, and G. Roger, Phys. Rev. Lett. {\bf 66},
1418
  (1991).

\bibitem{LaPSlu89}
A. La~Porta, R.~E. Slusher, and B. Yurke, Phys. Rev. Lett. {\bf
62},  28
  (1989).

\bibitem{PoiGra93}
J.~P. Poizat and P. Grangier, Phys. Rev. Lett. {\bf 70},  271
(1993).

\bibitem{RocVig97}
J.-F. Roch, K. Vigneron, P. Grelu, A. Sinatra, J.-P. Poizat, and
P. Grangier,
  Phys. Rev. Lett. {\bf 78},  634  (1997).

\bibitem{LevAbr93}
J.~A. Levenson, I. Abram, T. Rivera, P. Fayolle, J.~C. Garreau,
and P.
  Grangier, Phys. Rev. Lett. {\bf 70},  267  (1993).

\bibitem{PerOu94}
S.~F. Pereira, Z.~Y. Ou, and H.~J. Kimble, Phys. Rev. Lett. {\bf
72},  214
  (1994).

\bibitem{SchBru96}
S. Schiller, R. Bruckmeier, M. Schalke, K. Schneider, and J.
Mlynek, Europhys.
  Lett. {\bf 36},  361  (1996).

\bibitem{BruSch97a}
R. Bruckmeier, K. Schneider, S. Schiller, and J. Mlynek, Phys.
Rev. Lett. {\bf
  78},  1243  (1997).

\bibitem{BucAnd01}
B.~C. Buchler, P.~K. Lam, H.-A. Bachor, U.~L. Andersen, and T.~C.
Ralph, Phys.
  Rev. A {\bf 65},  011803(R)  (2001).

\bibitem{BenLev95}
K. Bencheikh, J.~A. Levenson, P. Grangier, and O. Lopez, Phys.
Rev. Lett. {\bf
  75},  3422  (1995).

\bibitem{BruHan97}
R. Bruckmeier, H. Hansen, and S. Schiller, Phys. Rev. Lett. {\bf
79},  1463
  (1997).

\bibitem{FriMuk00}
S.~R. Friberg, T. Mukai, and S. Machida, Phys. Rev. Lett. {\bf
84},  59
  (2000).

\bibitem{HauWat89}
H.~A. Haus, K. Watanabe, and Y. Yamamoto, J. Opt. Soc. Am. B {\bf
6},  1138
  (1989).

\bibitem{DruBre94}
P.~D. Drummond, J. Breslin, and R.~M. Shelby, Phys. Rev. Lett.
{\bf 73},  2837
  (1994).

\bibitem{CouSpa98}
J.~M. Courty, S. Sp\"alter, F. K\"onig, A. Sizmann, and G. Leuchs,
Phys. Rev. A
  {\bf 58},  1501  (1998).

\bibitem{KozIva02}
V.~V. Kozlov and D.~A. Ivanov, Phys. Rev. A {\bf 65},  023812
(2002).

\bibitem{GraLev98}
P. Grangier, J.~A. Levenson, and J.-P. Poizat, Nature {\bf 396},
537  (1998).

\bibitem{KonZie02}
F. K\"onig, M.~A. Zielonka, and A. Sizmann,
Phys. Rev. A
  {\bf 66},  013812  (2002).

\bibitem{SilLam01}
C. Silberhorn, P.~K. Lam, O. Wei\ss, F. K\"onig, N. Korolkova, and
G. Leuchs,
  Phys. Rev. Lett. {\bf 86},  4267  (2001).

\bibitem{RosShe91}
M. Rosenbluh and R.~M. Shelby, Phys. Rev. Lett. {\bf 66},  153
(1991).

\bibitem{SizLeu99}
A. Sizmann and G. Leuchs,  in {\em The optical Kerr effect and
quantum optics
  in fibers}, Vol.~XXXIV of {\em Progress in Optics}, edited by E. Wolf
  (Elsevier Publishers B. V., Amsterdam, 1999), p.\ 373.

\bibitem{Wer96}
M.~J. Werner, Phys. Rev. A {\bf 54},  R2567  (1996).

\bibitem{FriMac96}
S.~R. Friberg, S. Machida, M.~J. Werner, A. Levanon, and T. Mukai,
Phys. Rev.
  Lett. {\bf 77},  3775  (1996).

\bibitem{SpaKor98}
S. Sp\"alter, N. Korolkova, F. K\"onig, A. Sizmann, and G. Leuchs,
Phys. Rev.
  Lett. {\bf 81},  786  (1998).

\bibitem{Agr95}
G.~P. Agrawal, {\em Nonlinear Fiber Optics}, 2 ed. (Academic
Press, San Diego,
  1995).

\bibitem{MolEva91}
L.~F. Mollenauer, S.~G. Evangelides, and J.~P. Gordon, IEEE J.
Lightwave Tech.
  {\bf 9},  362  (1991).

\bibitem{Formel}
This formula is an extension of a calculation from
\cite{MolEva91}.

\bibitem{SpaBoh97}
S. Sp\"alter, M. B\"ohm, M. Burk, B. Mikulla, R. Fluck, I. Jung,
G. Zhang, U.
  Keller, S. A., and G. Leuchs, Appl. Phys. B {\bf 65},  335  (1997).

\bibitem{ChiWen89}
S. Chi and S. Wen, Opt. Lett. {\bf 14},  1216  (1989).

\bibitem{MecKum97}
A. Mecozzi and P. Kumar, Opt. Lett. {\bf 22},  1232  (1997).

\bibitem{LevVas99}
D. Levandovsky, M.~V. Vasilyev, and P. Kumar, Opt. Lett. {\bf 24},
43  (1999).

\bibitem{SchKno00}
E. Schmidt, L. Kn\"oll, D.-G. Welsch, M. Zielonka, F. K\"onig, and
A. Sizmann,
  Phys. Rev. Lett. {\bf 85},  3801  (2000).

\bibitem{LohGru94}
W.~H. Loh, A.~B. Grudinin, V.~V. Afanasjev, and D.~N. Payne, Opt.
Lett. {\bf
  19},  698  (1994).

\bibitem{BraPat01}
S.~L. Braunstein and A.~K. Pati, {\em Quantum Information Theory
with
  Continuous Variables} (Kluwer Academic, Dodrecht, 2001).

\bibitem{SouSch97}
P.~H. Souto~Ribeiro, C. Schwob, A. Ma\^{i}tre, and C. Fabre, Opt.
Lett. {\bf
  22},  1893  (1997).

\bibitem{SizKon00}
A. Sizmann, F. K\"onig, M.~A. Zielonka, R. Steidl, and T.
Rechtenwald,  in {\em
  Massive WDM and TDM Soliton Transmission Systems}, edited by A. Hasegawa
  (Kluwer Academic Publishers, Dodrecht, The Netherlands, 2000), p.\ 289.

\end{thebibliography}

\end{document}